\documentclass[aps,groupedaddres8s,fleqn,nofootinbib,twocolumn]{revtex4}
\usepackage{graphicx}
\usepackage{xcolor}
\usepackage{amsmath,amssymb}
\usepackage{float}
\usepackage{amsfonts}
\usepackage{verbatim}
\usepackage{flushend}
\usepackage{balance}
\usepackage{endnotes}
\usepackage{footnote}
\usepackage{adjustbox}
\usepackage{gensymb}
\usepackage{float}
\usepackage{epigraph}
\usepackage{xcolor}
\usepackage[utf8]{inputenc}
\usepackage{setspace}

\newcommand{\beq}{\begin{eqnarray}}
\newcommand{\eeq}{\end{eqnarray}}
\usepackage{amsmath}

\begin{document}

\title{Viscosity and diffusion in life processes and tuning of fundamental constants}
\author{K. Trachenko$^{1}$}
\address{$^1$ School of Physical and Chemical Sciences, Queen Mary University of London, Mile End Road, London, E1 4NS, UK}

\begin{abstract}
Viewed as one of the grandest questions in modern science, understanding fundamental physical constants has been discussed in high-energy particle physics, astronomy and cosmology. Here, I review how condensed matter and liquid physics gives new insights into fundamental constants and their tuning. This is based on two observations: first, cellular life and the existence of observers depend on viscosity and diffusion. Second, the lower bound on viscosity and upper bound on diffusion are set by fundamental constants, and I briefly review this result and related recent developments in liquid physics. I will subsequently show that bounds on viscosity, diffusion and the newly introduced fundamental velocity gradient in a biochemical machine can all be varied while keeping the fine-structure constant and the proton-to-electron mass ratio intact. This implies that it is possible to produce heavy elements in stars but have a viscous planet where all liquids have very high viscosity (for example that of tar or higher) and where life may not exist. Knowing the range of bio-friendly viscosity and diffusion, we will be able to calculate the range of fundamental constants which favor cellular life and observers and compare this tuning with that discussed in high-energy physics previously. This invites an inter-disciplinary research between condensed matter physics and life sciences, and I formulate several questions that life science can address. I finish with a conjecture of multiple tuning and an evolutionary mechanism.
\end{abstract}

\maketitle

\tableofcontents

\section{Introduction: fundamental constants and fine tuning}
\label{intro}

Our search for understanding, consistency and predictability of the physical world is based on mathematical structures \cite{tegmark}, our fundamental theories, which describe matter and fields. Our fundamental theories have about 20 fundamental physical constants such as the Planck constant $\hbar$, speed of light in vacuum $c$, the electron mass $m_e$ and charge $e$, the Newton constant and other parameters related to couplings setting the properties of elementary particles in the Standard Model \footnote{The values of fundamental physical constants are often listed in reviews and textbooks (see, e.g., Refs. \cite{uzanreview,uzan1,codata,ashcroft}), and their recommended values are maintained and updated in the National Institute for Standards and Technology database \cite{nist-fund}.}. These constants give the observed Universe its distinctive character and differentiate it from others we might imagine \cite{barrow,barrow1,carrbook,finebook,carr1,carr,cahnreview,hoganreview,adamsreview,uzanreview,uzan1}.

We do not know what kind of theories we need to explain the values of fundamental constants and their origins \cite{weinberg}. Due to these limitations, the fundamental constants are considered arbitrary \cite{cahnreview}. Some reports indicate that fundamental constants can in fact vary \cite{variable1,variable2}. Questioning the constancy of these constants amounts to questioning our fundamental theories such as the Standard model \cite{uzanreview}.

Understanding fundamental constants is considered to be among the grandest questions in modern science \cite{grandest}. Referring to fundamental constants as ``barcodes of ultimate reality'', Barrow adds that these constants will one day unlock the secrets of the Universe \cite{barrow}.

Fundamental constants govern a wide range of high-energy processes, starting from cosmology and inflation to nuclear reactions and nuclear synthesis in stars producing carbon, oxygen and other elements which can then form molecular structures essential to life. An interesting observation is that the values of some fundamental physical constants are finely tuned and balanced to give our observable world.

One example is the Hoyle's prediction of the energy level of carbon nucleus of 7.65 MeV. This resonance level is required in order to rationalise carbon abundance and in particular the synthesis of carbon from fusing three alpha particles in stars. After the Hoyle's prediction, the required energy level was found experimentally. This carbon resonance-level coincidence is viewed as striking. A related important effect is a slightly lower resonance level in oxygen, enabling carbon to survive other resonant reactions. This finely balanced set of coincidences enables carbon-based life. Production of carbon and oxygen depends on nuclear energy levels and the fine structure constant $\alpha=\frac{1}{4\pi\epsilon_0}\frac{e^2}{\hbar c}\approx\frac{1}{137}$ and strong nuclear force constant. According to our theories, a small change of these constants (more than 0.4\% and 4\% for the nuclear and fine structure constant) results in almost no carbon or oxygen produced in stars \cite{barrow,carrbook,finebook}. $\alpha$ and the proton-to-electron mass ratio $\beta=\frac{m_p}{m_e}\approx 1836$ also play a role in making stars hot enough to initiate and sustain nuclear reactions. Unless $\alpha$ and $\beta$ satisfy a certain relation, there would be no heavy nuclei made in stars. Unless $\alpha$ lies in a fairly narrow range, protons are predicted to decay long before the stars can form. There are other examples of what would happen were fundamental constants different, suggesting a fairly narrow ``habitable zone'' in the parameter space ($\alpha$,$\beta$) (see, however, Ref. \cite{adamsreview}). In this zone, stars are able to evolve and produce essential biochemical elements including carbon, life-supporting molecular structures can emerge and planets can form \cite{barrow,carrbook,finebook}. Another example is the finely-tuned balance between the masses of down and up quarks: larger down-quark mass gives the proton world without neutrons where light hydrogen atoms can form only but not heavy atoms, whereas larger up-quark mass gives the neutron world without protons and hence no atoms consisting of nuclei and electrons around them. Our world with heavy atoms and electrons endowing complex chemistry would disappear with a few per cent fractional change in the mass difference of the two quarks \cite{hoganbook,hoganreview,adamsreview}. For this reason, Barrow calls the observed fundamental constants ``bio-friendly'' \cite{barrow} and Adams refers to our Universe as ``biophilic'' \cite{adamsreview}.

Trying to rationalise fundamental constants, their balance and fine tuning has given rise to the anthropic principle (AP), sometimes referred to as the anthropic argument \cite{adamsreview,schellekens} or anthropic observation \cite{smolin}. Eliciting different views \cite{barrow,barrow1,vilenkin,hoganreview,adamsreview,uzanreview,carr,carr1,carrbook,smolin,finebook}, this term is a collection of related ways to rationalise the observed values of fundamental constants by proposing that these constants serve to create conditions for an observer to emerge and hence are not unexpected. Two popular formulations of the AP are the weak anthropic principle (WAP) and the strong anthropic principle (SAP). According to WAP, our expectations to observe must be restricted by the condition necessary for our presence as observers. WAP is a selection argument formulated in the cosmological context to explain why the Universe is so old and big: currently observed age of the Universe, its low density and temperature are consistent with the existence of an observer because it takes an age of a star to produce heavy elements necessary for observers to emerge. To explain the coincidences such as the Hoyle resonance and fine tuning of fundamental constants, the SAP states that the Universe and its fundamental constants must be such as to admit at some stage the emergence of observers \cite{barrow}.

To address the narrow range and fine tuning of fundamental constants and the anthropic principle, different proposals were discussed by eminent physicists \cite{carrbook}. The range of these proposals is large, extending to what some considered to be at the edge of science \cite{carrbook}. For example, the multiverse proposal requires an ensemble of universes and a physical mechanism to generate the ensemble with a certain distribution of fundamental constants in each universe. Then, a relatively small number of universes have the right values of fundamental constants, and we find ourselves in one of those universes and measure those constants. An alternative is the cosmological natural selection where bouncing black hole create new universes with fundamental constants close to those in the parent universe \cite{smolin}. Then, proliferating universes are those with fundamental constants conducive to the production of stars leading to black holes. We live in one of such proliferating universes where heavy nuclei required for observers are produced in stars.

Discussions of constraints on fundamental constants and their fine tuning involve high-energy processes at different scales and often end with production of heavy nuclei in stars. This involves a tacit assumption that once heavy nuclei are produced in stars, observers emerge. As discussed in the next section, there is a long way between nuclei and observers capable of building complex instruments such as accelerators and telescopes and interpreting the observations to ask the question about fundamental constants.

It is appreciated that what follows the creation of nuclei - processes involving the formation of proteins, RNA, living cells and so on - may be chemically and biologically complex. Some of the elements linking nuclei to observers were considered before. For example, Barrow and Tipler consider biochemistry and list essential properties of several chemicals and substances (e.g. carbon, carbon dioxide, carbonic acid, water, hydrogen, oxygen, nitrogen and its compounds) which are essential for the operation of biological molecules in living systems. The effects of fundamental constants on bond strengths, reaction energies and geometry of water and other simple molecules were studied, showing fairly moderate variations and indicating that the associated ``chemical'' constraints on fundamental constants are much less stringent as compared to those needed for the nuclear synthesis \cite{constantchem}. Ascertaining this for larger molecules important for life such as proteins and DNA is hard \cite{ellisuzan}, however a similar effect is expected. These discussions of relation between fundamental constants and chemistry are complemented by discussions of complexity and networks. These properties and emergent varieties are thought to contribute to enough complexity for life to emerge \cite{carr1}.

Due to their complexity and variety, these processes are not thought to be describable by a physical model which can relate them to physical constants and put bio-friendly constraints on these constants. 

What has been missing is a physical model which is both general enough to be widely applicable and yet specific enough to connect life processes directly to fundamental constants (as is done, for example, in fundamental theories predicting the bio-friendly values of $\alpha$, $\beta$, quark masses and so on). It may be hard to imagine that such a model would ever exist because life processes involved at the condensed matter scale are too complex and hence preclude a model which is both general enough and relatable all the way to fundamental constants.

Here, I will show that these models are nevertheless possible and will review how condensed matter physics can offer models which fill the large gap between heavy nuclei and observers. These models come from recent developments in liquids physics which I briefly review. These models are (a) general enough to be used on par with the fundamental theories considered previously, (b) are firmly ground in a wide range of experimental data and (c) impose constraints on fundamental constants which are different from those discussed before. These constraints come from bounds on viscosity, diffusion and the fundamental velocity gradient in a biochemical machine which I introduce. This implies that it is possible to produce heavy elements in stars but have a viscous planet where all liquids have very high viscosity and where life may not exist. Knowing the range of bio-friendly and observer-friendly viscosity and diffusion, we will be able to calculate the range of fundamental constants which favor cellular life and observers and compare this tuning with that discussed in high-energy and particle physics and astronomy previously. This invites an inter-disciplinary research between condensed matter physics and life sciences, and I formulate several questions that life scientists can address (e.g., is there a living-to-non-living transition at high viscosity or low diffusion constant?) I finish with a conjecture of multiple tuning and an evolutionary mechanism.

\section{Why condensed matter and liquids?}
\label{why}

As mentioned in the Introduction, nearly all discussions of our bio-friendly or biophilic Universe end at the production of heavy nuclei in stars. This involves a tacit assumption that once heavy nuclei form, observers emerge. While heavy nuclei are a necessary condition, it is clearly not a sufficient one. What happens between nuclei and observers asking questions about fundamental constants?

Let's first look at scales involved. Figure 1 shows different levels of structures in the physical world. These scales are similar to those used in the Cosmic Uroborus diagram \cite{carr1}. The distance between the size of nuclei and human observers, 15 orders of magnitude, is substantial and is not far from 20 orders of magnitude separating the Planck length from the nuclei.

\begin{figure*}
\begin{center}
{\scalebox{0.5}{\includegraphics{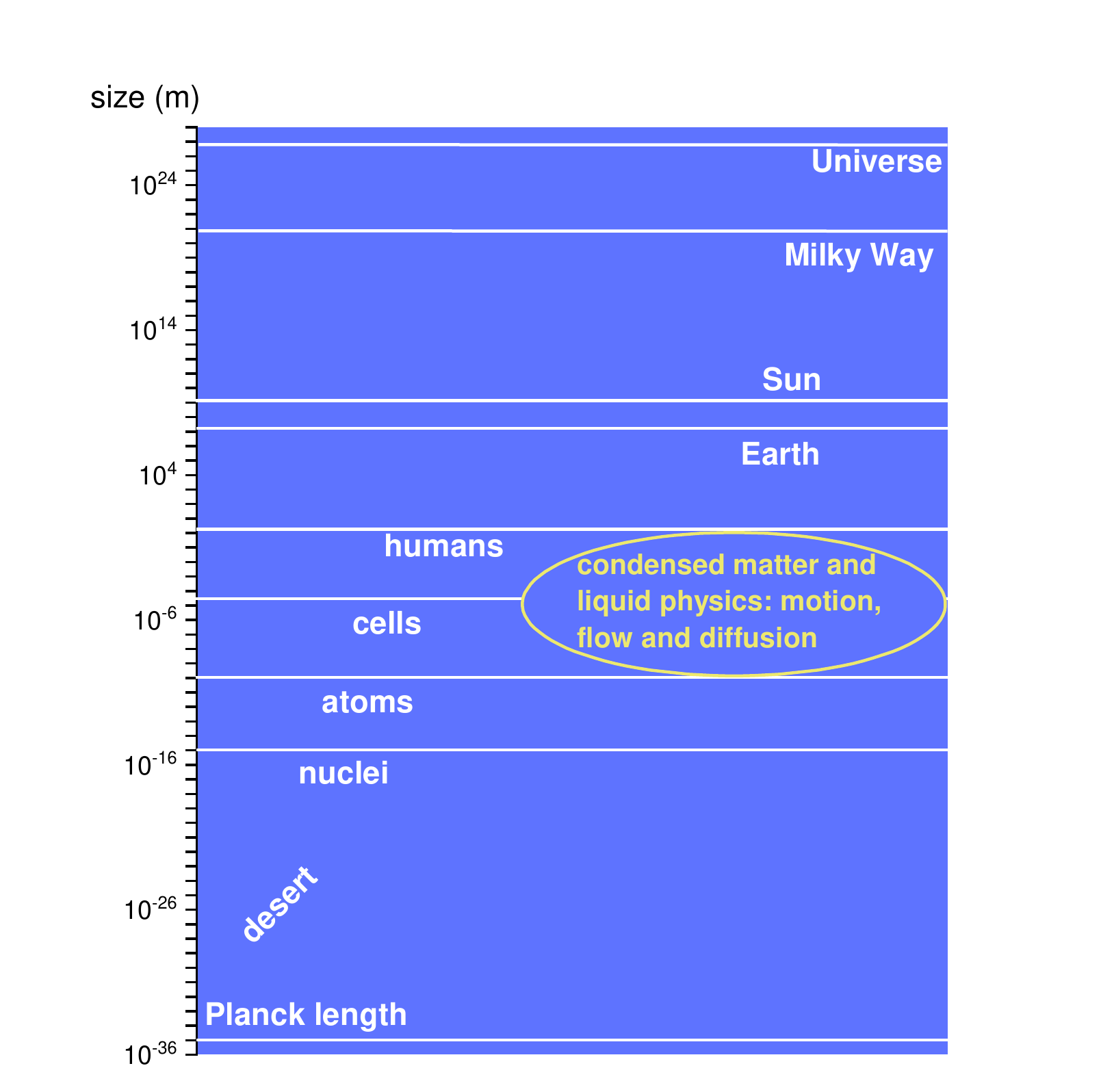}}}
\end{center}
\caption{Different scales of structure in the physical world.}
\label{visc}
\end{figure*}

The most important element leading to life and observers is the living cell, the basic structural and functional unit of life forms. Physical processes in condensed states of matter, solids and liquids, in the range roughly from micrometers to centimeters are described by condensed matter physics. With development of nanoscience, this range now approaches the atomistic scale on the short side. The long side is comparable with typical experimental samples in condensed matter research (e.g. in neutron or X-ray scattering experiments) in the mm-cm range. Hence, a physical model related to cells includes condensed matter physics. Inter-cellular processes and flow of liquids in the body can be described by continuum models such as fluid mechanics, however explaining and understanding parameters in the fluid mechanics such as viscosity and diffusion is still in the realm of condensed matter physics.

There are several areas related to cells where liquid flow is important. The two important ones are the operation of the cell itself (e.g., protein motors and cytoskeletal filaments, passive and active molecular transport, cytoplasmic mixing, mobility of cytoplasmic constituents, diffusion involved in cell proliferation and so on) and the flow processes in the entire body involving many cells such as blood flow \cite{biochembook,cellpaper1,cellpaper2,cellpaper3,cellpaper4}). Another area where the flow of matter is important is related to the pre-biotic synthesis of important molecules which cells use to operate, such as the metabolic flux. Here, the flow of gases feeds the Krebs cycle at the heart of life and creates life building blocks including amino acids, fatty acids, sugars and nucleotides \cite{lanebook}. Some of essential chemical reactions involved in these processes take place in liquid water, hence liquids come into play in the pre-biotic synthesis as well.

Liquids and gases are two states of matter providing a medium where this flow can happen and matter can {\it move}. Liquids in particular carry dense flow and can efficiently transport flux of matter essential to life. Viscosity governs this flow and is therefore tightly embedded in life processes and their dynamics.

In our world, the motion-enabling liquid is water, however the physical model discussed in Section \ref{minvisc} and its implications apply to all liquids. If life in a different world is not water-based but uses another liquid as a medium to provide motion (e.g., methane, carbon dioxide or liquid metal), the model implications are the same.

I will discuss the liquid state and its main properties involved in operation in cells, organisms and pre-biotic environments: dynamic viscosity $\eta$, kinematic viscosity $\nu=\frac{\eta}{\rho}$, where $\rho$ is density, and diffusion constant $D$. These properties govern flow and its velocity in stationary and time-dependent processes as well as diffusion. I will show how a physical model constrains the bounds of these properties in terms of fundamental physical constants. I will then discuss what would happen to life processes if these fundamental constants were different.

\section{Why now?}

This section is not essential for understanding results related to constraints on fundamental constants from bio-friendly viscosity and diffusion. Those interested in implications of these constraints for life processes can turn over to Section \ref{biofriendly}, having consulted Eqs. \eqref{nu1} and \eqref{nuf} showing fundamental lower bounds for viscosity.

This section serves to give a perspective into the problem of liquid theory and briefly reviews the history of liquid research. This is important since it shows the connection to more recent developments leading to fundamental bounds on viscosity. An emphasis is placed on experimental testability of this theory because this point is often brought up in the discussion of the anthropic principle and its alternatives \cite{carrbook}. I will show how a robust physical model for lower viscosity bound can be developed and directly compared to experimental data.

The discussion of the role of fundamental constants, their fine-tuning and the anhtropic principle has been limited to high-energy processes and models from particle physics, cosmology and astronomy. More recently, it was realised that fundamental constants also govern condensed matter properties and often in unexpected ways \cite{myreview}. This Section serves to show the path connecting a property such as viscosity and fundamental constants. An interesting point here is that numerical values of properties such as viscosity are hard to predict using a theoretical model, however there is a particular point on the phase diagram (e.g. minimum or maximum) where such a prediction is possible and, moreover, is made using fundamental physical constants only.

\subsection{Theory}
\label{theoryprob}

We have alluded to the importance of the liquid state for life processes. Among three basic states of matter (solid, liquid, gas), it is the liquid state which has fundamental theoretical problems. As discussed by Landau, Lifshitz and Pitaevskii (LLP), these problems combine (a) strong interatomic interactions combined with dynamical disorder and (b) the absence of a small parameter \cite{landaustat,pitaevskii,ahiezer}. For this reason, no general theory of liquids was long thought to be possible, in contrast to theories of solids and gases developed over a century ago. Whereas calculating generally-applicable thermodynamic properties such as energy and heat capacity and their temperature dependence are essential part of theories of solids and gases, deriving such general relations was considered impossible in liquids \cite{landaustat,pitaevskii}. According to Peierls \cite{peierls-frenkel}, Landau had always maintained that a theory of liquids is impossible overall.

The first part of the problem stated by LLP is illustrated by writing the liquid energy as

\begin{equation}
E=\frac{3}{2}NT+\frac{n}{2}\int g(r)u(r)dV
\label{enint}
\end{equation}

\noindent where $n$ is concentration, $u(r)$ is the interaction potential, $g(r)$ is the pair distribution function, interactions and correlations are assumed to be pairwise and $k_{\rm B}=1$.

Early liquid theories  \cite{kirkwoodbook,borngreen,zwanzig,barkerhenderson} considered that the goal of liquid theory is to provide a relation between liquid thermodynamics and liquid structure and intermolecular interactions such as $g(r)$ and $u(r)$ in Eq. \eqref{enint}. Working towards this goal involved developing the analytical models for liquid structure and interactions. This development has become the essence of theoretical approaches to liquids \cite{kirkwoodbook,borngreen,zwanzig,barkerhenderson,egelstaff,faber,march,marchtosi,tabor,balucani,hansen2,hansen1}. The problem is that the interaction $u(r)$ in liquids is both strong and system-specific, hence $E$ in Eq. \eqref{enint} is strongly system-dependent as stated by LLP. It was for this reason that no generally applicable theory of liquids was considered possible \cite{landaustat,pitaevskii}. An additional problem is that interatomic interactions and correlation functions are generally not available apart from fairly simple model liquids and can be generally complex involving many-body, hydrogen-bond interactions and so on. This precludes calculation of the liquid energy in the approach based on Eq. \eqref{enint} or its extensions involving, for example, higher-order correlation functions \cite{borngreen,barkerhenderson}. Even when $g(r)$ and $u(r)$ are available in simple cases, the calculation involving Eq. \eqref{enint} or similar is not enough: in order to explain experimental data such as temperature dependence of heat capacity of real liquids \cite{ropp,wallacecv,wallacebook,proctor1,proctor2,chen-review}, one still needs to develop a physical model in this approach.

Commonly-used liquid models are inapplicable to understanding the energy and heat capacity of real liquids. These models include the Van der Waals model, the hard-spheres model and their extensions \cite{hansen2,ziman,march,parisihard}. Both models give the specific heat $c_v=\frac{3}{2}k_{\rm B}$ \cite{landaustat,wallacecv,wallacebook}, the ideal-gas value, in contrast to experiments universally showing liquid $c_v=3k_{\rm B}$ close to melting \cite{ropp,wallacecv,wallacebook,proctor2}. These models were also used as reference states to calculate the energy \eqref{enint} by decomposing interactions into repulsive and attractive parts (see, e.g., Refs. \cite{barkerhenderson,wca1,wca2,zwanzig,rosentar}). These parts understandably play different roles at low and high density, however this method faces the same problem stated by LLP: the interactions and coefficients in the decomposition are strongly system-dependent and so are the final results, precluding a general theory.

In solids, both crystalline and amorphous, the above problems do not emerge because the solid state theory is based on collective excitations, phonons. This theory is physically transparent, predictive and applicable to all solids. There is no need to explicitly consider structure and interactions to understand basic thermodynamic properties of solids. Most important results such as the universal temperature dependence of energy and heat capacity of solids readily come out in the phonon approach \cite{landaustat}.

The small parameter in solids simplifying the solid state theory is small phonon displacements from equilibrium, but this seemingly does not apply to liquids because liquids do not have stable equilibrium points that can be used to sustain these small phonon displacements. Weakness of interactions assumed in the theory of gases does not apply to liquids either because interactions in liquids are as strong as in solids. This constitutes the second, no small parameter problem, stated by LLP.

As a result of these two fundamental problems, theoretical calculation and understanding energy and heat capacity of real classical liquids (both its values and temperature dependence) has remained a long-standing problem. This was noted as a notable gap in both research and undergraduate teaching \cite{chen-review,granato,prescod}.

It is interesting that although the approach to the liquid theory diverged from the solid state theory in its fundamental perspective, there were notable exceptions. Sommerfeld \cite{somm} and Brillouin \cite{br1,br2,br3,brillouin,brilprb} considered that the liquid energy and thermodynamic properties are fundamentally related to phonons as in solids and sought to discuss liquid properties on the basis of a modified Debye theory of solids. Just how this works in liquids was not clear at the time. The first Sommerfeld paper was published only 1 year after the Debye theory of solids \cite{debyepaper} and 6 years after the Einstein's paper ``Planck’s theory of radiation and the theory of the specific heat'' in 1907 \cite{einstein}. The problem of liquid thermodynamics and in particular its relationship to phonons has extended over a substantial period of Brillouin's research. Apart from isolated attempts \cite{wannier,faber,wallacebook}, this line of enquiry has stalled in the years that followed, and liquid theories based on structure and interactions were pursued instead. Whereas the Debye and Einstein theories have become part of nearly every textbook where solids and phonons are mentioned, a theory of liquid thermodynamics has remained unworkable for about a century that followed. One potential reason for this is that, differently from solids, the nature of phonons in liquids remained unclear for a long time. This is discussed in my book in detail \cite{mybook}.

Once looked from a wider and longer-term perspective, the history of liquid research reveals a fascinating story involving physics luminaries \cite{mybook}. In addition to the largely unknown line of enquiry by Sommerfeld and Brillouin, milestone contributions came from Maxwell in 1867 \cite{maxwell} and followed by Frenkel \cite{frenkel}. Frenkel wrote, but did not solve, the key equation to describe phonons in liquids \cite{frenkel,ropp}. This telegraph equation was referred to as the telegraphist equation by Poincar\'e \cite{poincare}, derived by Heaviside \cite{heaviside} in 1876 and earlier by Kirchhoff \cite{kirchhoff} in 1857. The Kirchhoff \index{Kirchhoff} paper precedes the Maxwell paper \cite{maxwell} by 10 years, and is probably the earliest publication that contains an equation relevant to understanding liquids.

Problems involved in liquid theory started to lift fairly recently and involved several stages \cite{mybook}. The first step involved the consideration of microscopic dynamics of liquid particles envisaged by Frenkel \cite{frenkel}: differently from solids where particle dynamics is purely oscillatory and gases where dynamics is purely diffusive/ballistic, particle dynamics in liquids is {\it mixed} and combines oscillations around quasi-equilibrium points as in solids and diffusive motions between different points.

The second step was ascertaining the nature of excitations in liquid using the microscopic dynamics above. At the fundamental level, physics of an interacting system is set by its excitations or quasiparticles \cite{landaustat,landaustat1}. In solids, these are phonons. The nature of phonons and their properties in liquids were not clear for a long time since Sommerfeld first discussed this problem in 1913 \cite{somm}. A fairly recent combination of theory, experiments and modelling led to understanding the propagation of phonons in liquids with an important attribute: the phase space available to these phonons is not fixed as in solids but is instead variable \cite{ropp,proctor1,proctor2,chen-review}. This is a non-perturbative effect and follows from the decay of the shear velocity field $v$ in liquids as \cite{ropp,yangprl,gapreview}


\begin{equation}
v\propto\exp\left(-\frac{t}{2\tau}\right)\exp\left(i(kx-\omega t)\right)
\label{gms4}
\end{equation}

\noindent with

\begin{equation}
\omega=\sqrt{c^2k^2-\frac{1}{4\tau^2}}
\label{gms5}
\end{equation}

\noindent where $\tau$ is liquid relaxation time, $c$ is the high-frequency transverse speed of sound, $k$ is the wavevector and $\omega$ is the frequency of transverse waves.

According to Eq. \eqref{gms5}, transverse waves exist in liquids if $k>k_g=\frac{1}{2c\tau}$ only. Consistent with molecular dynamics simulations \cite{yangprl}, this also implies that liquids are able to support shear stress at low frequency if system sizes are small enough. This has been ascertained experimentally \cite{noirez1,noirez2}.

According to Eq. (\ref{gms4}), the decay time (lifetime) \index{lifetime} and decay rate are $2\tau$ and $\Gamma=\frac{1}{2\tau}$. The crossover between propagating and non-propagating modes corresponds to $\omega=\Gamma$. Then, the propagating regime $\omega>\Gamma$ gives $k>\frac{1}{c\tau\sqrt{2}}=k_g\sqrt{2}$ or, using Eq. \eqref{gms5},

\begin{equation}
\omega>\frac{1}{2\tau}
\label{omprop}
\end{equation}

\noindent in agreement with the frequency (energy) gap envisaged originally by Frenkel \cite{frenkel}.

We see that the phonon phase space in liquids reduces with temperature: higher temperature reduces $\tau$ and increases $k_g$ above which transverse phonons exist as well as the frequency above which transverse phonons are propagating modes in \eqref{omprop}. This reduction has a general implication: the calculated specific heat of classical liquids universally decreases with temperature, as is seen experimentally \cite{ropp}. This theory based on the close relationship between liquid $c_v$ and the reduction of the phonon phase space has undergone an independent detailed and rigorous verification \cite{proctor1,proctor2}.

Considering the phonon phase space in liquids and its variation addresses the no small parameter problem stated by Landau, Lifshitz and Pitaevskii: the small parameter does exist in liquids but it operates in a reduced phase space. This is discussed in my book \cite{mybook} in detail.

The density of states in liquids is worth noting here, not least because it has been a source of confusion. The density of states $g(\omega)\propto\frac{k^2}{\frac{d\omega}{dk}}$ corresponding to the dispersion relation \eqref{gms5} is

\begin{equation}
g(\omega)\propto\omega^2\sqrt{1+\frac{1}{4\omega^2\tau^2}}
\label{gomega}
\end{equation}

In the solidlike elastic low-temperature regime $2\omega\tau\gg 1$ corresponding to the propagating regime \eqref{omprop}, $g(\omega)\propto\omega^2$. This is consistent with the well-established close similarity of experimental dispersion curves in solids and low-temperature liquids, including in the linear $\omega\propto k$ regime where $\omega$ is small \cite{ropp,copley,pilgrim,burkel,pilgrim2,water-tran,hoso,hoso3,monaco1,monaco2,sn}. This similarity readily implies the quadratic density of states $g(\omega)\propto\omega^2$ in both solids and low-temperature liquids. In the opposite non-propagating hydrodynamic high-temperature regime $2\omega\tau\ll 1$ ($\omega\ll\Gamma$), $g(\omega)\propto\frac{\omega}{\tau}$ (this is similar to the linear density of states of purely relaxing modes derived using a different approach \cite{zacconepnas}). $g(\omega)$ therefore crosses over from the quadratic Debye density of states in the propagating regime to the linear density of states in the non-propagating regime. This crossover is seen in the numerical instantaneous normal mode (INM) analysis \cite{keyes}. We therefore see that $g(\omega)$ in \eqref{gomega} originating from Eq. \eqref{gms5} explains both experimental dispersion curves in low-temperature liquids and the INM results, earlier and more recent \cite{keyes,rabani,shajin,egaminm}, including the quadratic and linear $g(\omega)$ as well as the crossover between them (particular types of disorder and relaxation may alter this picture \cite{zaccone-review}). Importantly, we see that the liquid energy due to propagating phonons is related to the quadratic Debye density of states, in agreement with experiments related to phonons \cite{copley,pilgrim,burkel,pilgrim2,water-tran,hoso,hoso3,monaco1,monaco2,sn,ropp,mybook} and independent analysis of thermodynamic implications for liquid properties \cite{proctor1,proctor2}.


In relation to liquids, the quadratic density of states $g(\omega)\propto\omega^2$ was first used by Migdal and Landau \cite{landaustat,landaumigdal} to calculate the specific heat, $c_v$, of liquid $^4$He, resulting in $c_v\propto T^3$ as in solids. This is probably the earliest calculation relating phonons and thermodynamics in a liquid, classical or quantum, which withstood the test of time and experiment. Both $c_v\propto T^3$ and the calculated magnitude of $c_v$ itself are in close agreement with high-precision experiments in liquid $^4$He (see, e.g., Refs. \cite{greywall,wilksbook}). This gives another clean evidence that $g(\omega$) of propagating phonons in liquids is quadratic rather than not: for example, the linear density of states would have given $c_v\propto T^2$ not experimentally observed in liquid $^4$He.

The above theory of liquids related to the variable phase space is closely based on details of microscopic particle dynamics. Considering this microscopic dynamics also turns out to be the key to understanding the lower bound of liquid viscosity and relating this bound to fundamental physical constants. This is a recent result \cite{sciadv1} and is discussed in the next section.

\subsection{Lower bound on liquid viscosity in terms of fundamental physical constants}
\label{minvisc}

Ascertaining the lower bound on liquid viscosity involves two steps: (a) understanding the origin of viscosity minima and (b) relating these minima to fundamental physical constants.

\begin{figure}
\begin{center}
{\scalebox{0.4}{\includegraphics{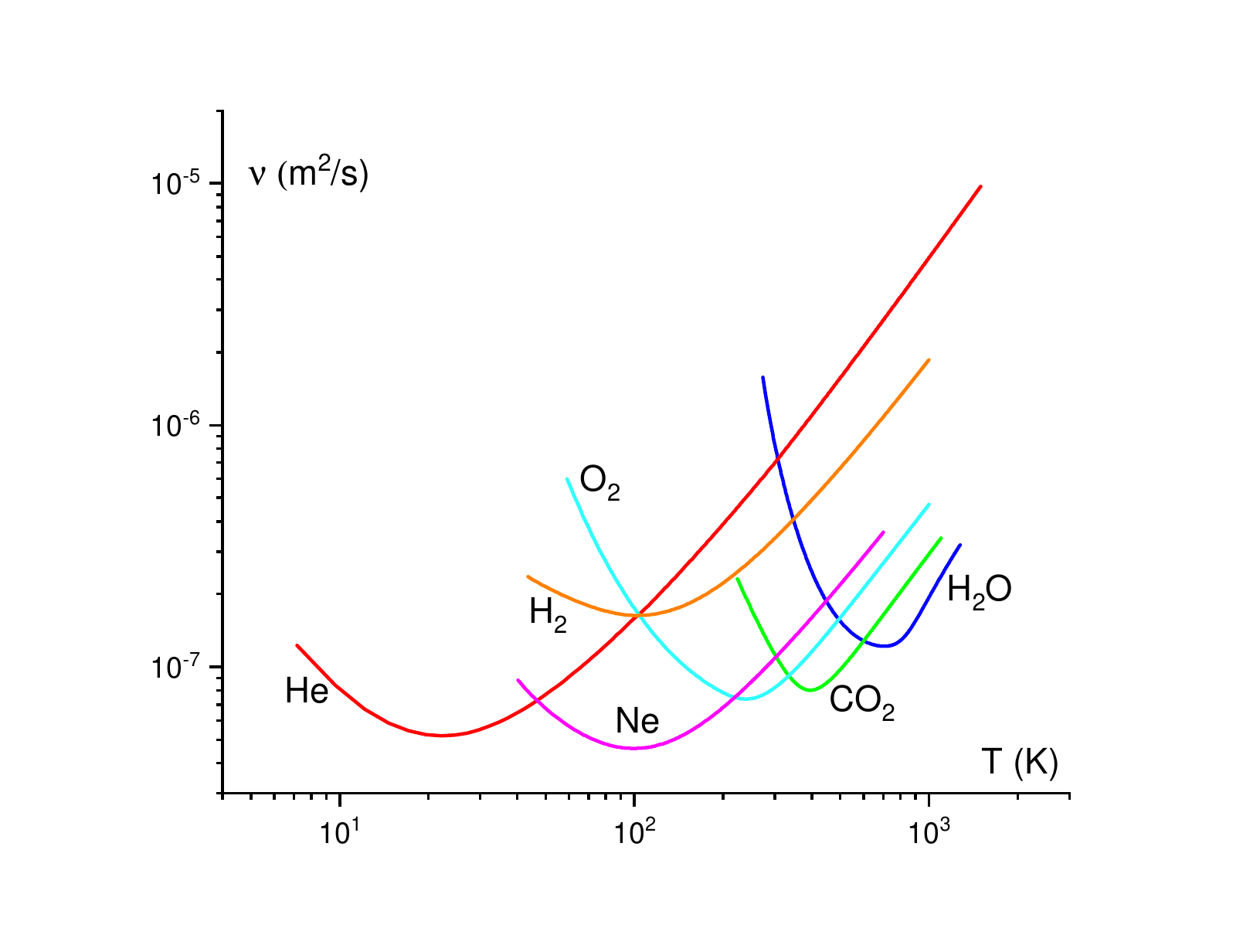}}}
\end{center}
\caption{Experimental kinematic viscosity of noble, molecular and network fluids showing minima. $\nu$ for He, H$_2$, O$_2$, Ne, CO$_2$ and H$_2$O are shown at 20 MPa, 50 MPa, 30 MPa, 50 MPa, 30 MPa and 100 MPa, respectively. The experimental data are from Ref. \cite{nist}.
}
\label{minima}
\end{figure}

Figure \ref{minima} shows experimental kinematic viscosity of several types of liquids. The viscosity shows minima, the universal effect in all liquids \cite{nist}.

Viscosity minima can be understood by zeroing in on the microscopic dynamics of particles. In a dilute gas, viscosity $\eta$ is set by molecules flying in straight lines up to the mean free path distance $L$ and transferring momentum during collisions:

\begin{equation}
\eta=\frac{1}{3}\rho v L
\label{gaslike}
\end{equation}

\noindent where $\rho$ and $v$ are density and average velocity of molecules, respectively.

Eq. \eqref{gaslike} predicts that $\eta$ of gases does not depend on density $\rho$ or pressure because $L=\frac{1}{n\sigma}$ and $\rho=nm$, where $n$ is concentration, $\sigma$ is molecule cross-section area and $m$ is molecule mass. It also predicts that gas viscosity {\it increases} with temperature because $v$ increases. This increase might be counter-intuitive because we often experience fluids getting thinner when heated: think about bitumen flowing in hot weather or jam getting thicker when cooled in the fridge after cooking.

Differently from gases, viscosity of dense fluids is set by the molecules vibrating around quasi-equilibrium positions before jumping to the neighbouring sites. The frequency of these jumps increases with temperature, and viscosity decreases with temperature as a result:

\begin{equation}
\eta=\eta_0\exp\left(\frac{U}{k_{\rm B}T}\right)
\label{liquidlike}
\end{equation}

\noindent where $U$ is activation energy and $\eta_0$ is the pre-factor setting the limiting high-temperature viscosity \cite{frenkel} which can be experimentally measured (see, e.g., Refs \cite{nussinov,nussinov2}).

The increase of $\eta$ at high temperature and its decrease at low imply that $\eta$ has a {\it minimum}, and we expect this minimum to correspond to the crossover between the gas-like and liquid-like viscosity. This crossover is related to the dynamical crossover at the Frenkel line separating liquidlike and gaslike dynamics \cite{flreview}.

It is convenient to look at this viscosity crossover above the critical point where it is smooth and no liquid-gas phase transition intervenes as in Figure \ref{minima}. Below the critical point, viscosity still has minima, albeit involving sharper changes. Viscosity values at the minima are close above and below the critical point \cite{sciadv1}.

Calculating viscosity analytically is hard, for the same reason stated by LLP: interactions in liquids are strong and system-specific, and so is $U$ in Eq. \eqref{liquidlike}. However it turns out that the minimum of viscosity at the crossover represents a very special point where viscosity can in fact be evaluated, even though approximately. More specifically, it turns out that the minima of kinematic viscosity $\nu$, $\nu_{min}$, can be related to only two basic properties of a condensed matter system: interatomic separation $a$ and the largest frequency in the system close to the Debye frequency $\omega_{\rm D}$ \cite{sciadv1}:

\begin{equation}
\nu_{min}=\frac{1}{2\pi}\omega_{\rm D}a^2
\label{nu}
\end{equation}

These two parameters, $a$ and $\omega_{\rm D}$, are related to the ``UV'' cutoff in a condensed matter system. They can be related to the Bohr radius, $a_{\rm B}$, setting the characteristic scale of inter-particle separation on the order of Angstrom as

\begin{equation}
a_{\rm B}=\frac{4\pi\epsilon_0\hbar^2}{m_e e^2}
\label{bohr}
\end{equation}

\noindent and the Rydberg energy, $E_{\rm R}=\frac{e^2}{8\pi\epsilon_0a_{\rm B}}$ \cite{ashcroft}, setting the characteristic scale for the cohesive energy in condensed matter phases on the order of several eV:

\begin{equation}
E_{\rm R}=\frac{m_ee^4}{32\pi^2\epsilon_0^2\hbar^2}
\label{rydberg}
\end{equation}

\noindent where $e$ and $m_e$ are electron charge and mass.

We note that the cohesive energy, or the characteristic energy of electromagnetic interaction, is

\begin{equation}
E=\frac{\hbar^2}{2m_ea^2}
\label{direct}
\end{equation}

This energy becomes $E_{\rm R}$ in Eq. \eqref{rydberg} if $a=a_{\rm B}$ is used. Relation \eqref{direct} will become useful later when we discuss the variation of viscosity with the Planck constant $\hbar$.

Relating $\omega_{\rm D}$ in Eq. \eqref{nu} to the cohesive energy $E$ using the known ratio

\begin{equation}
\frac{\hbar\omega_{\rm D}}{E}=\left(\frac{m_e}{m}\right)^{\frac{1}{2}}
\label{ratio}
\end{equation}

\noindent and setting $a$ and $E$ to their characteristic scales $a_{\rm B}$ and $E_{\rm R}$ gives

\begin{equation}
\nu_{min}=\frac{1}{4\pi}\frac{\hbar}{\sqrt{m_em}}
\label{nu1}
\end{equation}

\noindent where $m$ is the molecule mass \cite{sciadv1}.

Two fundamental constants, $\hbar$ and $m_e$, appear in Eq. \eqref{nu1}. The minimal viscosity turns out to be quantum. This may seem surprising and at odds with our thinking about high-temperature liquids as classical. Eq. \eqref{nu1} reminds us that the nature of interactions in condensed matter is quantum-mechanical in origin, with $\hbar$ affecting the Bohr radius and Rydberg energy. The minimal viscosity does not depend on the electron charge which, while present in Eq. \eqref{bohr} and \eqref{rydberg}, cancels out in Eq. \eqref{nu1}.

\subsection{Relation to experiments}

We started this paper with discussing fine-tuning of fundamental constants. This fine-tuning follows from physical models. For example, changing the balance between the quark masses results in either the neutron world without protons or the proton world without neutrons and heavy atoms. We can not check this result experimentally. Instead, this follows from our fundamental theories of particle physics which we compare to experiments in other ways. Verifying these theories experimentally is a challenge, and this point is at the centre of the collection of ingenious papers related to fine-tuning and the anthropic principle \cite{carrbook}. Whereas verifying condensed matter theories should be easier, this can be a challenge as well. For this reason, we expand on this point here more generally.

Quoting Popper, Smolin observes \cite{smolin} that ``a theory is falsifiable if one can derive from it unambiguous predictions for practical experiments, such that - were contrary results seen - at least the premise of the theory would have been proven not to true''.
The need for a theory to be falsifiable and make testable experimental predictions is of course not new. We have learned this from the success of physics built from such models: our most important theories forming the core of modern physics were born out of closely following actual processes in real systems and experimental data \cite{weinberg1}. Reviewing the history of early and modern science, Weinberg observes that ``the world acts on us like a teaching machine'', shaping our theories and reliable knowledge. We may think about different theories but we learn about the world by keeping only those theories agreeing with experimental data.

Compared to cosmology, astronomy and particle physics, condensed matter has more opportunities to test a theory experimentally: we can study liquids under different lab conditions, chemical compositions and use a large variety of probes, from a simple dropping ball bench experiment measuring viscosity to neutron scattering requiring large research facilities. Although these possibilities are ample, the focus of liquid theories discussed in Section \ref{theoryprob} historically has often been on liquid models rather than experimental data. This was probably related to attempts to overcome theoretical problems set out by LLP in Section \ref{theoryprob}. As a result, it is not uncommon to find papers and books on liquids, glasses, disordered systems and spin glasses where the focus ends up mainly on models rather than comparing theory to experiments. This includes areas where authors state upfront that they choose not to discuss theory predictions and comparisons with real world and experimental data (see, for example, pp. xi-xii of Ref. \cite{parisi}).

An apt and practical view was taken by Landau and Peierls \cite{landaupeierls}:

\begin{quotation}
``The essence of any physical theory is to use the results of an experiment to make assertions about the results of a future experiment.''
\end{quotation}

The emphasis here is on experimental result and relations between them, the view also held by Bohr \cite{mermin}. Eq. \eqref{nu1} is an example: if we can experimentally measure fundamental constants featuring in this equation, the equation predicts the smallest value of viscosity a liquid can ever get.

For different fluids such as those in Fig. \ref{minima}, Eq. (\ref{nu1}) predicts $\nu_m$ in the range (0.3-1.5)$\cdot 10^{-7}\frac{{\rm m}^2}{\rm s}$. The ratio between experimental and predicted $\nu_m$ is in the range of about 0.5-3 \cite{sciadv1}.


Another and more general way to compare Eq. \eqref{nu1} to experiments is to write the mass as

\begin{equation}
m=Am_p
\label{mass}
\end{equation}

\noindent where $m_p$ is the proton mass and $A$ is the number of protons and neutrons and define a ``fundamental viscosity'' by setting $A=1$ corresponding to H (similarly to (\ref{bohr}) and (\ref{rydberg}) derived for the H atom).

\begin{equation}
\nu_f=\frac{1}{4\pi}\frac{\hbar}{\sqrt{m_em_p}}\approx 10^{-7}~ {\rm \frac{\rm{m}^2}{\rm{s}}}
\label{nuf}
\end{equation}

Eq. \eqref{nuf} differs from Eq. \eqref{nu1} by a factor $\frac{1}{\sqrt{A}}$. For many liquids such as those shown in Figure \ref{minima}, this factor is not too large to alter the viscosity minima substantially.

\begin{figure}
\begin{center}
{\scalebox{0.25}{\includegraphics{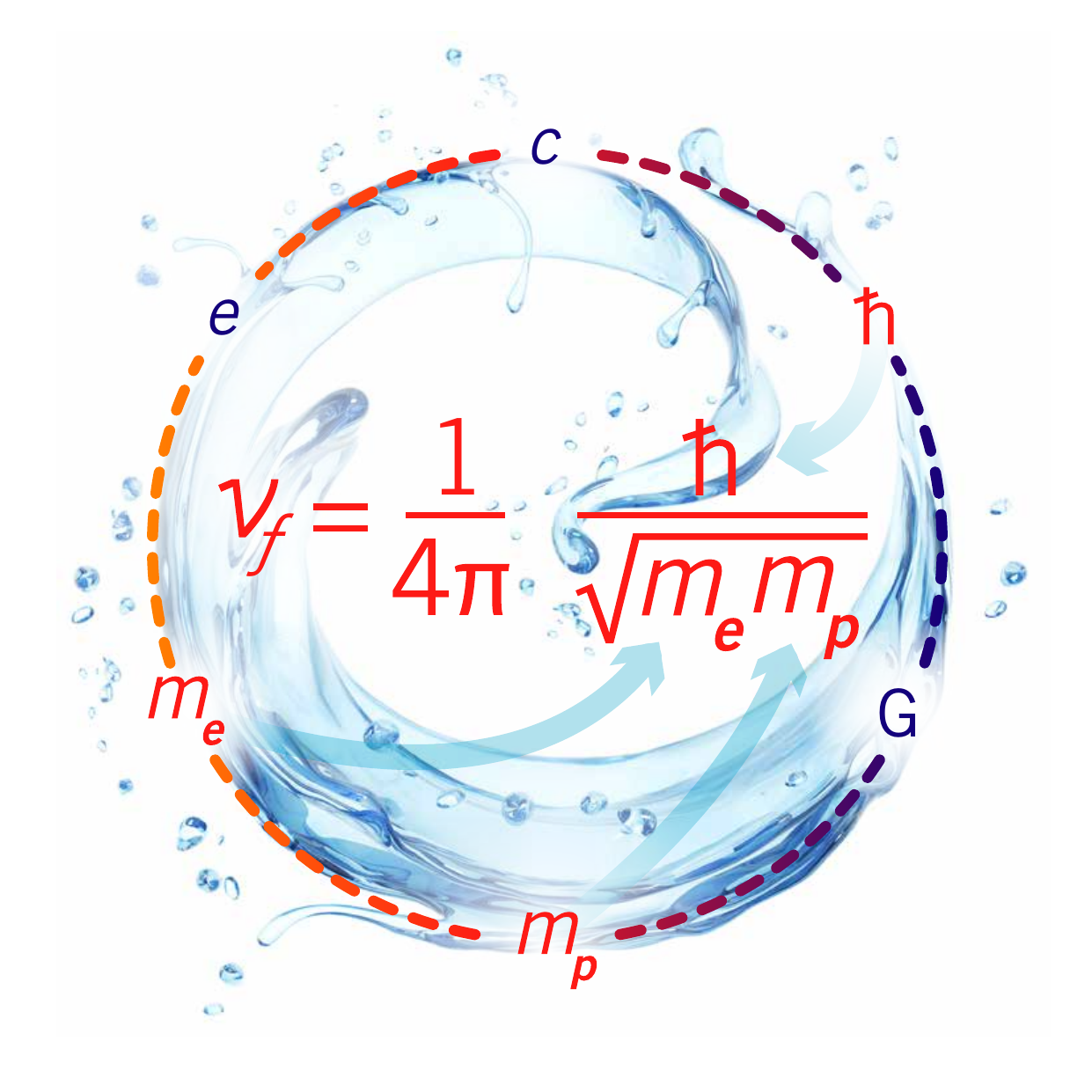}}}
\end{center}
\caption{Fundamental kinematic viscosity $\nu_f$ depends on three values only: $\hbar$, $m_e$ and $m_p$.
}
\label{fundamental}
\end{figure}

As illustrated in Figure \ref{fundamental}, the fundamental minimal quantum viscosity depends on three constants: $\hbar$, $m_e$ and $m_p$ (although $m_p$ depends on other parameters in the Standard Model, it is often listed as a fundamental constant \cite{uzanreview,uzan1,codata,ashcroft,nist-fund}; the proton-to-electron mass ratio is discussed as the important finely-tuned dimensionless constant \cite{barrow,adamsreview}).

The fundamental minimal viscosity $\nu_f\approx 10^{-7}~ {\rm \frac{\rm{m}^2}{\rm{s}}}$ agrees well with experimental results in Fig. \ref{minima}.

More recently, the experimental viscosity of metallic liquids was discussed at high temperature in order to find limiting high-temperature viscosity and found it close to the prediction of Eq. \eqref{nu1} \cite{nussinov,nussinov1,nussinov2}. 

The upshot of the above discussion of viscosity minima is that we have a physical model for liquid viscosity which is verifiable experimentally. This is important in view of the emphasis on experimental testability of models involved in discussions of fine-tuning and anthropic principle \cite{carrbook}.

Interestingly, the model of minimal viscosity answers the long-standing question regarding experimental viscosity considered by Purcell and Weisskopf. Nearly 50 years ago, Purcell observed that there is almost no liquid with viscosity much lower than that of water and wrote in the first paragraph of Ref. \cite{purcell} (original italics preserved):\\

``The viscosities have a big range {\it but they stop at the same place}. I don't understand that''.\\

In the first footnote of that paper, Purcell says that Weisskopf has explained this to him. I did not find published Weisskopf's explanation, however the same year Weisskopf published the paper ``About liquids'' \cite{weisskopf}. That paper starts with a story often recited by conference speakers: imagine a group of isolated theoretical physicists trying to deduce the states of matter using quantum mechanics only. They are able to predict the existence of gases and solids, but not liquids. Weisskopf implies that liquids are hard, and we listed the main reason for this in Section \ref{theoryprob}.

We now understand the reason for the effect that puzzled Purcell: liquid viscosities stop decreasing because they have minima, and those minima are fixed by fundamental constants. These minima happen to be not far from viscosity of water at room conditions.

Finally, we recall the emphasis of Smolin on experimental testability of a physical model used to discuss fine tuning and anthropic principle. This section shows that condensed matter physics offers plenty of possibilities to test a model, once this model becomes available. This puts the implications of the model on a solid basis.

\section{Bio-friendly viscosity and diffusion: implications for life processes}
\label{biofriendly}

\subsection{Viscosity, relaxation time and diffusion constant in terms of fundamental constants}
\label{limits}

We observe life processes in different life forms and can therefore discuss factors that enable these processes. We saw that fundamental physical constants set the lower bound of liquid viscosity. The cause and effect here are fundamental constants and viscosity bound, respectively. We now flip the cause and effect and ask what happens to liquid viscosity and life processes that depend on this viscosity when fundamental constants are different? In particular, what constraints are imposed on the fundamental constants from essential life processes in and between cells where motion and flow are involved \cite{sciadv2023}?

The minimal viscosity changes in response to fundamental constants according to Eq. \eqref{nu1}. If the minimum increases towards bio-unfriendly values, viscosity necessarily becomes larger at {\it all} conditions of pressure and temperature, in {\it all} liquids (not just in water essential in our world).

Let's consider this process in more detail and write the Navier-Stokes equation as

\begin{equation}
\rho\frac{\partial {\bf v}}{\partial t}=-\nabla p+\eta\nabla^2{\bf v}
\label{navier}
\end{equation}

\noindent where ${\bf v}$ is the fluid velocity which is assumed to be small and $p$ is pressure.

For time-dependent flow, the solution of Eq. \eqref{navier} depends on kinematic viscosity $\frac{\eta}{\rho}$. We first consider steady flow where the flow velocity depends on $\eta$. Using $\eta_m=\nu_m\rho$, $\rho\propto\frac{m}{a_{\rm B}^3}$, $m\propto m_p$ and Eqs. \eqref{bohr} and \eqref{nu1}, we find

\begin{equation}
\eta_{min}\propto\frac{e^6}{\hbar^5}\sqrt{m_pm_e^5}
\label{etamin}
\end{equation}

A useful dynamical quantity related to $\eta_{min}$ is the minimal liquid relaxation time

\begin{equation}
\tau_{min}=\frac{\eta_{min}}{G}
\label{tau1}
\end{equation}

\noindent where $G$ is the high-frequency shear modulus.

Liquid relaxation time $\tau$ corresponds to the average time it takes a molecule to jump from one quasi-equilibrium place to the next \cite{frenkel}. In terms of fundamental constants, the upper bound to elastic moduli in condensed phases can be estimated as \cite{myreview}

\begin{equation}
G\propto\frac{E_{\rm R}}{a_{\rm B}^3}
\label{tau2}
\end{equation}

Combining Eqs. \eqref{tau1}, \eqref{tau2} and $\eta_{min}=\nu_{min}\frac{m}{a_{\rm B}^3}$ as before gives

\begin{equation}
\tau_{min}=\frac{\nu_{min}m}{E_{\rm R}}
\label{tau21}
\end{equation}

Using Eqs. \eqref{rydberg}, \eqref{nu1} and \eqref{mass} with $A=1$ as before, Eq. \eqref{tau21} gives

\begin{equation}
\tau_{min}\propto\frac{\hbar^3}{m_ee^4}\left(\frac{m_p}{m_e}\right)^{\frac{1}{2}}
\label{tau3}
\end{equation}

We obtain the same value $\tau_{min}$ in terms of fundamental constants if we note that this time is related to the shortest time scale in the system set by the Debye vibration period, $\tau_{\rm D}$. Writing $\tau_{\rm D}=\frac{1}{\omega_{\rm D}}$, using $\omega_{\rm D}$ from Eq. \eqref{ratio}, $E=E_{\rm R}$ from Eq. \eqref{rydberg} and Eq. \eqref{mass} with $A=1$ as before, we find $\tau_{\rm D}=\tau_{min}$ up to a constant factor.

$\eta_{min}$ in Eq. \eqref{etamin} and $\tau_{min}$ in Eq. \eqref{tau3} depend on fundamental constants differently: for example, smaller $\hbar$ increases $\eta_{min}$ but decreases $\tau_{min}$. The physical origin of this difference is as follows. Smaller $\hbar$ gives larger $E_{\rm R}$ in Eq. \eqref{rydberg}, increasing the bond energy and bond stiffness. This increases $\omega_{\rm D}$ and decreases the Debye vibration period $\tau_{\rm D}$ as well as $\tau_{min}\propto\tau_{\rm D}$. Therefore, $\tau_{min}$ sets the short-time dynamics which becomes faster in response to the variation of fundamental constants which increases $E_{\rm R}$. On the other hand, viscosity and its minimal value $\eta_{min}$ increase with $E_{\rm R}$ and with the variation of fundamental constants causing this increase of $E_{\rm R}$. For example, smaller $\hbar$ increases both $E_{\rm R}$ in Eq. \eqref{rydberg} and $\eta_{min}$ in Eq. \eqref{etamin}.

We now address the diffusion properties. Using the Stokes-Einstein relation between $\eta$ and diffusion constant $D$ in the low-temperature liquidlike dynamical regime,

\begin{equation}
D=\frac{k_{\rm B}T}{6\pi r\eta}
\end{equation}

\noindent where $r$ is the radius of moving particle, we find the maximal $D$ corresponding to the minimal $\eta_m$ as

\begin{equation}
D_{max}\propto\frac{1}{\eta_{m}}\propto\frac{\hbar^5}{e^6}\frac{1}{\sqrt{m_pm_e^5}}
\label{dmax}
\end{equation}

Let's consider what happens if we dial $\hbar$ and set it smaller than the current value or increase $e$ so that the viscosity minimum \eqref{etamin} gets higher. $\eta_m$ in Eq. \eqref{etamin} is quite sensitive to $\hbar$ and $e$. Raising the viscosity minimum implies that viscosity increases at all conditions of pressure and temperature. Higher viscosity means that water now flows slower, dramatically affecting vital processes in cells. Large viscosity increase (think of viscosity of tar and higher) means that life might not exist in its current form or not exist at all. Examples of processes where this is the case are abound, and we only consider some of them.

Viscosity of cytoplasm in cells is about 1 Pa$\cdot$s \cite{isabel-cell} and is about 1000 times larger than water at room conditions due to the interactions related to cytoskeletons, organelles, machineries and other structures. This complex viscosity nevertheless increases with water viscosity and would disable essential processes if water viscosity becomes too high. Higher viscosity means that water flows slower, dramatically affecting vital flow processes in and between cells and so on. This includes the slowdown of essential chemical reactions involved in life processes such proteins folding and enzyme kinetics \cite{chemrate1,chemrate2,chemrate3}. Higher viscosity also slows down flow processes under an external gradient including active transport, and an example of this flow is discussed in section \ref{bmachine}. Viscosity increase similarly affects inter-cellular processes such as blood flow. Consider, for example, blood flow. Blood viscosity is considered normal in the range of about 3.5 cP to 5.5 cP, whereas moving significantly outside this range is not conducive to body functioning. The decrease of $\hbar$ in Eq. \eqref{etamin} by only 9\% increases $\eta_{min}$ by a factor of 1.6 (and decreases $D_{max}$ in Eq. \eqref{dmax}), covering the normal range and precluding significantly larger variations of $\hbar$.

Once might ask if viscosity increase due to different fundamental constants may be part of the overall slowing down (similar to a video slow motion) whereby all processes slow down but remain functioning. Several observations can be made in this regard. First, larger viscosity not only slows down dynamics but can arrest a life process. Examples include a transition corresponding to the explosive increase of the coagulation rate in biological fluids such as protein solutions and blood. This takes place at the critical value of the P\'eclet number which depends on viscosity \cite{zaccone-peclet}. This example represents a wider class of {\it extrinsic} factors affecting viscosity due to many-body collective effects in complex fluids such as, for example, blood (see, e.g., Refs. \cite{succi,bloodvisc,noirezblood1,noirezblood2}). These effects act on top of the intrinsic ``bare'' factors setting viscosity such as the energy and length scale in condensed matter which we considered in Eqs. \eqref{nu}, \eqref{bohr} and \eqref{rydberg}. Using appropriate rheology models, the extrinsic factors can be accounted for in the experimental data to arrive at the intrinsic viscosity to be used to evaluate the constrains on fundamental constants. This will be discussed in detail elsewhere.

Second, $\eta_{min}$ in Eq. \eqref{etamin} increases with $e$, $m_e$ and decreases with $\hbar$, whereas the elementary time $\tau_{min}$ in Eq. \eqref{tau3}, or the shortest time $\tau_{\rm D}$, do the opposite. As $\eta_{min}$ and viscosity increase due to, for example, larger $e$ or $m_e$ or smaller $\hbar$, $\tau_{min}$ decreases. This implies that in terms of the shortest atomic timescale $\tau_{min}$ ($\tau_{\rm D}$), time effectively runs faster, and processes dependent on short-time dynamics speed up rather than slow down.

Third, the chemical reaction rates of vital biological processes involving, for example, dynamics of proteins and enzymes, $k$, depend on $\eta$ as $k\propto\frac{1}{\eta^n}$, where $n$ varies in quite a large range: from 0.3 \cite{chemrate1} to 2.4 depending on the reaction (see, e.g., Ref. \cite{chemrate3} for review). Therefore, viscosity increase affects different reaction rates differently and disrupts the existing balance between products of different reactions and important interactions between those products. Depending on the nature and degree of this disruption, the result can either be finding a new functioning sustainable balance during life development and hence a different type of life or not finding a sustainable living state at all.

Viscosity increase leads to the decrease of diffusion constant $D$, according to Eq. \eqref{dmax}. This slows down all diffusive processes of essential substances and molecular structures in and across cells involving passive and facilitated transport \cite{biochembook} and affects molecular transport, cytoplasmic mixing, mobility of cytoplasmic constituents and sets the limits at which molecular interactions and biological reactions can occur. Diffusion is also essential for cell proliferation. Viscous and diffusive processes in and between cells have been of interest have been of interest in life, biomedical and biochemical sciences (see examples in Refs. \cite{cellpaper1,cellpaper2,cellpaper3,cellpaper4}).

Physically, the origin of viscosity increase at smaller $\hbar$ or larger $e$ is related to the decrease of the Bohr radius \eqref{bohr} as the classical regime with smaller $\hbar$ is approached. This results in the increase of the cohesive energy in Eq. \eqref{rydberg} via Eq. \eqref{direct}, making it harder to flow and diffuse because a flow event requires overcoming the energy barrier set by the cohesive energy. The same applies to the variation of other fundamental constants: for example, larger $e$ increases the cohesive energy \eqref{rydberg} and the minimal viscosity $\eta_{min}$ in Eq. \eqref{etamin}.

Large viscosity increase (think of viscosity of tar and higher) means that life might not exist in its current form or not exist at all. One might hope that cells could still survive in such a Universe by finding a hotter place where overly viscous and bio-unfriendly water is thinned. This would not help though: $\eta_m$ sets the minimum below which viscosity can not fall regardless of temperature or pressure as is seen in Fig. \ref{minima}. This applies to {\it any} liquid and not just water and therefore to all life forms using the liquid state to function.


We therefore see that water and life are well attuned to the values of fundamental constants including the degree of quantumness of the physical world set by $\hbar$. This applies to $e$, $m_e$ in Eqs. \eqref{etamin} and \eqref{dmax} and to a smaller degree to $m_p$.

\subsection{Constraints on fundamental constants from bio-friendly viscosity and diffusion}

Let $\eta_0$ be viscosity above which a life process is disabled (we will elaborate on this point in the next section) and $D_0$ be the diffusion constant below which life is impossible. Conditions for these two properties to be bio-friendly are

\begin{equation}
\eta_{min}<\eta_0
\label{etacond}
\end{equation}

\noindent and

\begin{equation}
D_0<D_{max}
\label{dcond}
\end{equation}

Combining Eqs. \eqref{etacond} and \eqref{dcond} with Eqs. \eqref{etamin} and \eqref{dmax}, we write

\begin{eqnarray}
\begin{split}
& \hbar>{\rm max}\left(\frac{1}{\eta_0^\frac{1}{5}}, D_0^{\frac{1}{5}}\right)\\
& m_e<{\rm min}\left(\eta_0^{\frac{2}{5}},\frac{1}{D_0^\frac{2}{5}}\right)\\
& m_p<{\rm min}\left(\eta_0^2,\frac{1}{D_0^2}\right)\\
& e<{\rm min}\left(\eta_0^{\frac{1}{6}},\frac{1}{D_0^\frac{1}{6}}\right)\\
\end{split}
\label{condi1}
\end{eqnarray}

I add two remarks regarding \eqref{condi1}. First, I have dropped the numerical and unit conversion factors which can be reinstated using previous equations, for the purpose of \eqref{condi1} is to show the trend of how a fundamental constant is constrained by viscosity and diffusion (assuming other constants are unchanged). This trend shows that in order for viscosity and diffusion to remain bio-friendly, the degree of quantumness, $\hbar$ should be large enough, whereas $m_e$, $m_p$ and $e$ should be small enough.

Second, the conditions for $\eta_{min}$ and $D_{max}$ are independent because they come from different life processes and can therefore disable them independently. The condition for $\eta_{min}$ is related to liquid flow in cells and organisms under an external gradient including active transport, and an example of this flow is discussed in section \ref{bmachine}. The condition for $D_{max}$ is related to essential diffusive processes such as passive and facilitated transport \cite{biochembook}. Depending on how sensitive biological and biochemical processes sensitive are to changes of flow and diffusion, these two conditions can be different. Hence I used the maximum for the constraint on $\hbar$ and the minimum for constraints on $m_e$, $m_p$ and $e$ so that the range \eqref{condi1} reflects the mechanism which disables a life process first.

In this section, I have considered viscosity getting too high and bio-unfriendly due to different fundamental constants increasing the lower viscosity bound. We could also consider changing fundamental constants in a ways that reduce viscosity by reducing the lower bound. If flow and diffusion become too fast due to lower viscosity, the accumulation of chemicals in cells and organisms might become too large for healthy functions. However, healthy viscosity and diffusion can be recovered by cells finding different external conditions: for example lower temperature or higher pressure would increase viscosity and decrease diffusion and flow back to their healthy levels as needed. Hence decreasing the lower viscosity bound is not as arresting for life as increasing the lower bound.

\subsection{Non-equilibrium flow}

For time-dependent non-equilibrium flow such as pulsed blood flow where kinematic viscosity $\nu=\frac{\eta}{\rho}$ plays a role in Eq. \eqref{navier}, the two conditions \eqref{etacond} and \eqref{dcond} need to be complemented by

\begin{equation}
\nu_{min}<\nu_0
\label{nucond}
\end{equation}

\noindent where $\nu_0$ is kinematic viscosity above which life processes are disabled.

Substituting $\nu_0$ in \eqref{nu1} by its fundamental value \eqref{nuf} adds additional constraints on $\hbar$, $m_e$ and $m_p$. Combining these constraints with those in Eq. \eqref{condi1}, we find

\begin{eqnarray}
\begin{split}
& {\rm max}\left(\frac{1}{\eta_0^\frac{1}{5}}, D_0^{\frac{1}{5}}\right)<\hbar<\nu_0\\
& \nu_0^2<m_e<{\rm min}\left(\eta_0^{\frac{2}{5}},\frac{1}{D_0^\frac{2}{5}}\right)\\
& \nu_0^2<m_p<{\rm min}\left(\eta_0^2,\frac{1}{D_0^2}\right)\\
& e<{\rm min}\left(\eta_0^{\frac{1}{6}},\frac{1}{D_0^\frac{1}{6}}\right)\\
\end{split}
\label{condi2}
\end{eqnarray}

\noindent and observe that bio-friendly constraints on $\eta$, $D$ as well as $\nu$ imply a bio-friendly {\it window} for $\hbar$, $m_e$ and $m_p$.

The window emerges because $\eta_{min}$ in Eqs. \eqref{nu1} and $\nu_{min}$ in \eqref{etamin} depend on $\hbar$, $m_e$ and $m_p$ differently.

\subsection{General constraints on fundamental constants}

As mentioned earlier, inequalities such as \eqref{condi1} and \eqref{condi2} impose constraints on each fundamental constant provided other constants do not change. A more general case is when more than one constant varies. In this case, combining \eqref{nu1}, \eqref{etamin} and \eqref{dmax} with \eqref{etacond}, \eqref{dcond} and \eqref{nucond}, we write the general set of constraints from bio-friendly $\eta$, $D$ and $\nu$:

\begin{eqnarray}
\begin{split}
& \frac{e^6}{\hbar^5}\sqrt{m_pm_e^5}<\eta_0\\
& \frac{\hbar^5}{e^6}\frac{1}{\sqrt{m_pm_e^5}}>D_0\\
& \frac{\hbar}{\sqrt{m_em_p}}<\nu_0
\end{split}
\label{condi3}
\end{eqnarray}

The first and second inequalities are trivially related. Combining one of them with the third inequality gives a more compact form. For example, multiplying the first and third inequality and using Eq. \eqref{bohr} gives

\begin{eqnarray}
\begin{split}
& a_{\rm B}>a^*\\
& a^*=\frac{1}{\sqrt{(4\pi)^3\epsilon_0}}\frac{e}{\sqrt{\eta_0\nu_0}}
\end{split}
\label{condi4}
\end{eqnarray}

\noindent where we re-instated numerical factors and $\epsilon_0$.

$a^*$ in Eq. \eqref{condi4} is the length setting the lower bound for the Bohr radius from the bio-friendly viscosity. Smaller range of bio-friendly viscosity set by $\eta_0$ and $\nu_0$ increase $a^*$, decreasing the range in which the Bohr radius remains bio-friendly.

\subsection{Variations of bounds on viscosity and diffusion at fixed $\alpha$ and $\beta$}
\label{alphabeta}

We have considered how viscosity, flow and diffusion can change in response to varying fundamental constants. This variation should be constrained because it should avoid the range where the production of lower-level structure is disabled. An important effect discussed in Section \ref{intro} is that the fine-structure constant

\begin{equation}
\alpha=\frac{1}{4\pi\epsilon_0}\frac{e^2}{\hbar c}
\label{al}
\end{equation}

\noindent and the proton-to-electron mass ratio

\begin{equation}
\beta=\frac{m_p}{m_e}
\label{be}
\end{equation}

\noindent need to be finely-tuned in order for heavy nuclei to be produced in stars.

In what follows, we fix $\alpha$ and $\beta$ as a way to reflect their fine tuning and ask how this affect viscosity, diffusion and flow.

One might think that the constraints on fundamental constants such as $\alpha$ and $\beta$ from star formation or nuclear synthesis are already tight enough to keep water viscosity from taking unwanted values not conducive to life. However, it is possible to substantially change the lower bounds for kinematic and dynamic viscosity and at the same time keep these constants intact, with no consequences for star formation and nuclear synthesis. There are many ways of doing this. For example, increasing $m_e$ and $m_p$ proportionally increases $\eta_m$ in Eq. \eqref{etamin} but leaves $\alpha$ and $\beta$ intact. Or, one can change $\hbar$ and $e$ in Eq. \eqref{etamin} but keep $\alpha$ and $\beta$ the same. To make this more explicit, one way to write $\eta_m$ in Eq. \eqref{etamin}, $D_{max}$ in Eq. \eqref{dmax} and $\nu_{min}\propto\nu_f$ in Eq. \eqref{nuf} in terms of $\alpha$ and $\beta$ in \eqref{al} and \eqref{be} is

\begin{eqnarray}
\begin{split}
&\eta_{min}\propto\left(\frac{e^2}{\hbar c}\right)^3\sqrt{\frac{m_p}{m_e}}\frac{m_e^3c^3}{\hbar^2}\\
&D_{max}\propto\frac{1}{\left(\frac{e^2}{\hbar c}\right)^3\sqrt{\frac{m_p}{m_e}}}\frac{\hbar^2}{m_e^3c^3}\\
&\nu_{min}\propto\frac{1}{\frac{e^2}{\hbar c}\sqrt{\frac{m_p}{m_e}}}\frac{e^2}{m_ec}
\end{split}
\label{fixed}
\end{eqnarray}

We note that the above bounds derived in non-relativistic condensed matter physics are not expressible in terms of dimensionless $\alpha$ and $\beta$ only, as is often the case for other bounds and properties in relativistic high-energy physics \cite{barrow,barrow1,carrbook}, but depend on other fundamental constants too.

We observe that fixing $\alpha$ and $\beta$ still leaves many ways of varying $\eta_{min}$, $D_{max}$ and $\nu_{min}$ in \eqref{fixed}. For example, any change of $\hbar$, $m_e$ or $c$ in the factor $\frac{m_e^3c^3}{\hbar^2}$ in $\eta_{min}$ and $D_{max}$ changes $\eta_{min}$ and $D_{max}$ but this change can always be compensated by changing other constants in $\alpha$ and $\beta$ to keep $\alpha$ and $\beta$ intact. This can be done in many ways. Similarly, changing $m_e$ in the factor $\frac{e^2}{m_ec}$ in $\nu_{min}$ in \eqref{fixed} changes $\nu_{min}$ but can be compensated by $m_p$ to keep $\beta$ intact. Or, changing $e$ in the factor $\frac{e^2}{m_ec}$ changes $\nu_{min}$ but can be compensated by the change of $\hbar$ in $\alpha\propto\frac{e^2}{\hbar c}$ and so on.

If $\eta_{min}$ is written as

\begin{equation}
\eta_{min}\propto\left(\frac{e^2}{\hbar c}\right)^3\sqrt{\frac{m_p}{m_e}}\left(\frac{m_ec}{\hbar}\right)^3\hbar
\label{etamin11}
\end{equation}

\noindent we observe that the inverse of $\frac{m_ec}{\hbar}$ is the reduced Compton wavelength of the electron, $\lambda_{\rm C}$. Although the Compton wavelength is physically unrelated to viscosity, Eq. \eqref{etamin11} shows that the lower viscosity bound can still vary even if we incorporate constraints from quantum and relativistic physics involved in Compton scattering and fix $\lambda_{\rm C}$, together with fixing $\alpha$ and $\beta$.

We therefore see that a Universe with fundamental constants different to ours can produce heavy elements in stars but have a planet where all liquids have very high viscosity due to large $\eta_{min}$ in \eqref{fixed}, for example that of tar or higher and where observers may not emerge. This can be achieved, for example, by increasing $m_e$ and/or decreasing $\hbar$ while keeping $\alpha$ and $\beta$ constant in \eqref{fixed} as discussed above. In order to reduce this high life-disabling $\eta_{min}$ to its current bio-friendly value, we need to dial the fundamental constants back to their current values so that the bounds \eqref{fixed} become bio-friendly. Hence we need to tune the {\it same} fundamental constants setting $\alpha$ and $\beta$ ($\hbar$, $e$, $c$, $m_e$, $m_p$) which, importantly, involves tuning that is additional and different to tuning involved in fixing $\alpha$ and $\beta$. This additional tuning due to bio-friendly viscosity is not needed for the generation of heavy nuclei and is therefore {\it redundant} for heavy nuclei. This redundancy involves vast differences between the two processes (generating heavy nuclei and bio-friendly viscosity and diffusion in living organisms) extending up to 15 orders of magnitude in size and similarly large energy difference as discussed in Section \ref{why}.

I note that this redundancy applies if tight constraints on $\alpha$ are relaxed \cite{adamsreview}. In this more general case, the constraints on the same fundamental constants from bio-friendly viscosity and diffusion in \eqref{fixed} are still additional and different from those imposed by the production of heavy nuclei.

I also note that viscosity, bio-friendly to motion in and between cells due to currently observed fundamental constants, plays a role even before heavy elements are formed in stars. For example, flow plays an important role in formation of stars from molecular clouds. Turbulence of this flow, which depends on viscosity, promotes star formation \cite{turbulence}. Higher viscosity due to different fundamental constants would reduce this turbulence, affecting star formation.

These observations bear a relation to questions asked previously: can we understand the values of fundamental constants on the basis of a theory more fundamental than we currently have (the Standard Model) \cite{barrow,barrow1,carrbook,finebook,weinberg}? How were these constants tuned \cite{carrbook,finebook}? One possibility is that fundamental constants were tuned once. As mentioned earlier, this would involve redundancy. Apart from redundancy itself, this simultaneous tuning would be somewhat different from how we view our physical theories: our current understanding is that each level of ``structure'' at a given energy and length scale in the Universe is described by an effective theory which has an underpinning theory of substructure at the next smaller length scale (apart from quarks and leptons), and each level of structure requires a different branch of physical theory \cite{carr}. Yet the above redundancy involves tuning processes and theories operating at vastly different length and energy scales.

If redundancy is to be avoided, we can conjecture that multiple independent tunings were involved. I develop this point in the Appendix.

I note that I have no stake in the debate of weak vs strong anthropic principle \cite{barrow,carrbook,finebook}, however the proponents of SAP can probably take the viscous planet with no observers as a supporting argument in a sense that in order for observers to emerge, the same fundamental constants need to be tuned at different levels. As mentioned earlier, these levels can be vastly separated and involve 15 orders of magnitude of length scale. I also note that I arrived at the conclusion of different tunings and redundancy for reasons unrelated to the need for observers to exist as posited in SAP. Rather, I (a) observed that the current values of fundamental constants are conducive to bio-friendly viscosity and diffusion involved in essential life processes and (b) combined this observation with the result that viscosity and diffusion have bounds set by fundamental constants.

\subsection{Flow in a biochemical machine}
\label{bmachine}

To complete the discussion of the role of fundamental constants in life processes involving motion and flow, I now derive the fundamental velocity gradient that can be set up in biochemical machines (molecular, cellular, inter-cellular or other). These machines play a vital role in sustaining cells and life.

Let's consider a machine creating an external force acting to move the liquid in or between cells. There is a limit to how efficient these machines are because they are powered by chemical energy, the energy of chemical bonds with a characteristic scale set by Eq. \eqref{rydberg}.

Let's consider this process in more detail and consider a liquid flowing with constant speed $u$ in direction $x$ in a volume $V$. The viscous stress is $\sigma_x=\eta\frac{\partial v}{\partial y}$, where $y$ is perpendicular to $x$ and $v$ linearly changes with $y$ as $v=\frac{yu}{l}$ in a simple planar geometry, where $l$ is distance between planes \cite{landaufluid}. The viscous force is then $f_x=\eta\frac{u}{l}S$, where $S$ is the area across our volume. The work done by this viscous force to move the liquid distance $x$ is $A=\eta\frac{u}{l}Sx=\eta\frac{u}{l}V$. The energy to do this work comes from released chemical, cohesive, energy $E$ (e.g., in the Krebs cycle in the metabolic flux) so we write $\frac{E}{V}=\eta\frac{u}{l}$. $E$ can be written as $NE_0$, where $N$ is the number of energy-releasing centres in a chemical network and $E_0$ is the cohesive energy in one bond whose order of magnitude is roughly given by $E_{\rm R}$ in Eq. \eqref{rydberg}. $V$ can be written as $NV_0$, where $V_0$ is the elementary volume approximately given by $a_{\rm B}^3$. This gives

\begin{equation}
\eta\frac{u}{l}=C\frac{E_{\rm R}}{a_{\rm B}^3}
\label{machine}
\end{equation}

\noindent where the coefficient $C$ absorbs different factors such as the density of energy-releasing centres, their energy and size in relation to $E_{\rm R}$ and $a_{\rm B}$, geometry of the molecular or cellular machine and so on.

$C$ is expected to be $C\ll 1$ since $E_{\rm R}$ is larger than the typical energy released in one event in the metabolic flux and $a_{\rm B}$ is smaller than the typical size of the energy-releasing centre. As a result, the volume density of energy, $\frac{E_{\rm R}}{a_{\rm B}^3}$ in Eq. \eqref{machine}, is larger than in a biochemical machine.

Eq. \eqref{machine} relates viscosity and the velocity gradient $\frac{u}{l}$ that can be set up in a liquid by a machine powered by chemical energy. $\frac{E_{\rm R}}{a_{\rm B}^3}$ in Eq. \eqref{machine} is fixed by fundamental physical constants and gives the maximal efficiency of a molecular or cellular machine to move the viscous liquid.

If the operating volume is roughly the same in all directions, $u$ can be assumed to be $u=\frac{l}{t_0}$ and the gradient $\frac{u}{l}$ becomes $\frac{1}{t_0}$, where $t_0$ is the characteristic transport time, the time it takes to move the flowing liquid in the operating volume across its own length. In this case, Eq. \eqref{machine} becomes

\begin{equation}
\eta=C\frac{E_{\rm R}}{a_{\rm B}^3}t_0
\label{machine1}
\end{equation}

We now recall the lower bound for viscosity discussed earlier, $\eta_{min}<\eta$. Combining this inequality with Eq. \eqref{machine} gives

\begin{equation}
\eta_{min}\frac{u}{l}<C\frac{E_{\rm R}}{a_{\rm B}^3}
\label{machine2}
\end{equation}

Writing $\eta_m=\nu_m\rho$, $\rho=\frac{m}{a_{\rm B}^3}$, $m=Am_p$ and using Eq. \eqref{nu1} as before, the inequality \eqref{machine2} becomes

\begin{equation}
\frac{u}{l}<\frac{C}{8\pi\sqrt{A}}\frac{{m_e}e^4}{\hbar^3\epsilon_0^2}\left(\frac{m_e}{m_p}\right)^{\frac{1}{2}}
\label{machine3}
\end{equation}

Eq. \eqref{machine3} gives the upper bound for the velocity gradient that can be set up by a biochemical machine powered by the chemical bond energy in terms of fundamental physical constants:

\begin{equation}
\left(\frac{u}{l}\right)_{max}=\frac{C}{8\pi\sqrt{A}\epsilon_0^2}\frac{{m_e}e^4}{\hbar^3}\left(\frac{m_e}{m_p}\right)^{\frac{1}{2}}
\label{machine31}
\end{equation}

In view of Eq. \eqref{rydberg} and \eqref{ratio}, Eq. \eqref{machine31} can be written as

\begin{equation}
\left(\frac{u}{l}\right)_{max}=4\pi C\omega_{\rm D}
\label{machine52}
\end{equation}

\noindent which makes good physical sense: the velocity gradient is limited by the largest frequency of atomic motion available in the condensed matter system, $\omega_{\rm D}$.

Using Eq. \eqref{machine31}, we introduce  the fundamental velocity gradient \cite{sciadv2023}:

\begin{equation}
\left(\frac{u}{l}\right)_f\propto\frac{{m_e}e^4}{\hbar^3}\left(\frac{m_e}{m_p}\right)^{\frac{1}{2}}
\label{machine4}
\end{equation}

In terms of $\alpha$ and $\beta$, the fundamental gradient can be written as

\begin{equation}
\left(\frac{u}{l}\right)_f\propto\frac{\left(\frac{e^2}{\hbar c}\right)^2}{\sqrt{\frac{m_p}{m_e}}}\frac{m_ec^2}{\hbar}
\label{machine5}
\end{equation}

\noindent where, interestingly, the factor $\frac{m_ec^2}{\hbar}$ is the Compton frequency of the electron.

Similarly to fundamental viscosity and diffusion discussed in Section \ref{alphabeta}, the fundamental gradient can be varied in ways which keep $\alpha$ and $\beta$ intact. Differently from $\eta_{min}$ in Eq. \eqref{etamin11} in Section \ref{alphabeta}, fixing the Compton frequency (together with $\alpha$ and $\beta$) fixes the fundamental gradient.

The inverse of the fundamental gradient $\left(\frac{u}{l}\right)_f$ is the fundamental transport time $t_f$:

\begin{equation}
t_f\propto\frac{\hbar^3}{{m_e}e^4}\left(\frac{m_p}{m_e}\right)^{\frac{1}{2}}
\label{machine8}
\end{equation}

\noindent which bounds the transport time $t_0$ from below:

\begin{equation}
t_0>t_f
\label{machine7}
\end{equation}

In addition to introducing new dynamical flow quantities, the above equations such as Eq. \eqref{machine5} make the point similar to that in the previous section: along with viscosity and diffusion, other flow properties essential for life such as the fundamental gradient can vary in response to variation of fundamental constants in such as a way as to keep $\alpha$ and $\beta$ intact and hence not to disturb the synthesis of heavy nuclei in stars. I will revisit this point in the Appendix.

\section{Inter-disciplinary questions}
\label{questions}

In Section \ref{biofriendly}, I have written bounds on fundamental constants that follow from bio-friendly viscosity and diffusion (Eqs. \eqref{condi1},\eqref{condi2}). This is probably as far as we can go using physics only. To calculate the actual bounds we need to know $\eta_0$, $D_0$ and $\nu_0$ in Eqs. \eqref{etacond}, \eqref{dcond}, \eqref{nucond} and accompanying inequalities in Section \ref{biofriendly}. This invites an input from the areas of biochemistry and biology and, more broadly, life sciences. A tentative list of questions includes:

\begin{enumerate}
\item What increase of viscosity and decrease of diffusion constant can life-enabling processes (including pre-biotic synthesis, operation of cells and inter-cellular processes) tolerate at each stage of life development?\\
\item Is the result of viscosity increase to slow down the function/process only? Is there a living-to-non-living {\it transition} at high viscosity $\eta$ (see, e.g., Ref. \cite{zaccone-peclet}) or low diffusion constant $D$ at different stages of life development?\\
\item Which processes are most sensitive to changes of viscosity and diffusion and are disabled first at each stage?
\item Provided we identify these processes, what are the consequences for other functions which depend on those processes? Can we draw a chart of inter-dependent biological and biochemical processes which depend on viscosity and diffusion?\\
\item If we can discuss these effects for carbon-based life forms, can we envisage other life forms where these effects may be different?
\label{questions}
\end{enumerate}

Questions 1-2 are related to the question of whether larger viscosity simply slows life processes down or whether these processes get disabled at some very high-viscosity (think of tar viscosity and higher). Our experience with cells and organisms based on water may seem to be at odds with the possibility that life can still operate if water viscosity became as high as tar viscosity due to different fundamental constants different. As mentioned in Section \ref{limits}, larger viscosity does not only slow down a function but can arrest a process altogether, as is the case in biological fluids such as protein solutions and blood \cite{zaccone-peclet}. The same applies to acceptable variations of water viscosity affecting cellular processes such as the diffusion of essential elements and molecular structures in and between cells. As discussed in Section \ref{limits}, larger viscosity also upsets the balance between different chemical reactions involving proteins, enzymes and so on and interactions between reaction products.

Life sciences may already have enough data to analyse in order to answer some of the questions above, or perhaps more data are needed. Once we calculate the bounds on fundamental constants from bio-friendly viscosity and diffusion, we can compare the range of allowed values to the range set by particle physics and synthesis of heavy nuclei \cite{barrow,adamsreview}. In turn, this will contribute to the discussion of the anthropic principle discussed in Section \ref{intro}. This will be also important for understanding the values of how fundamental constants and how these constants were tuned.

Regardless of implications for fundamental constants and fine tuning, the above questions above are probably interesting on their own in life sciences.

\section{Summary}

Interesting science is often at the interface \cite{wheeler}. In this overview, I pointed out that condensed matter physics has an important role to play in the discussion of fine tuning of fundamental physical constants and anthropic arguments discussed in high-energy physics. This is based on two insights: first, cellular life and the existence of observers depend on viscosity and diffusion. Second, the lower bound on viscosity and upper bound on diffusion are set by fundamental constants. Importantly, we saw that viscosity, diffusion and the newly introduced fundamental velocity gradient in a biochemical machine can all be varied while keeping the fine-structure constant and the proton-to-electron mass ratio intact. This implies that it is possible to produce heavy elements in stars but have a viscous planet where all liquids have very high viscosity, for example that of tar or higher, and where life may not exist.

Knowing the range of bio-friendly viscosity and diffusion, we can calculate the range of fundamental constants which favor cellular life and observers. This knowledge will come from biochemistry and biology, inviting another inter-disciplinary research between condensed matter physics and life sciences. Calculating the bio-friendly and observer-friendly range of fundamental constants from viscosity and diffusion, we will be able to compare this tuning with that discussed previously in high-energy particle physics and astronomy.

Regardless of implications for fundamental constants, thinking about the role of viscosity and diffusion in life processes prompts generally interesting questions for life sciences such as whether there is a living-to-non-living transition as a function of viscosity and diffusion?

We saw that the additional tuning due to bio-friendly viscosity is not needed for the generation of heavy nuclei and is therefore redundant for heavy nuclei. This redundancy involves large differences between the two processes (generating heavy nuclei and bio-friendly viscosity and diffusion in living organisms) in terms of energy and size, differences extending up to 15 orders of magnitude. If redundancy is to be avoided, we can conjecture that multiple independent tunings were involved. This point is developed in the Appendix where an evolutionary mechanism is discussed.

I am grateful to F. Adams, S. Arseniyadis, V. V. Brazhkin, B. Carr, R. Goldstein, S. de Kort, L. Noirez, N. Ojkic, Isabel M. Palacios, A. E. Phillips, S. Succi, U. Windberger and A. Zaccone for discussions and the EPSRC for support.

\section{Appendix: tuning and evolution}
\label{multiple}

As mentioned in the previous section, interesting science is often at the interface between different areas. The interface discussed in this review is well-defined and is between condensed matter physics and life sciences including biology and biochemistry. The questions related to this interface are well-posed. For example, given that (a) the lower viscosity bound and related properties are set by fundamental constants and (b) key life processes in and across cells are governed by viscous flow, what are the constraints on fundamental constants from bio-friendly viscosity and diffusion?

In this Appendix, we extend the connection between biology and recent insights from condensed matter physics to more fundamental areas exploring the origin of fundamental constants. This exploration is related to some of the grandest challenges in modern science \cite{grandest}. Given the scale of this challenge \cite{weinberg}, our discussion is predictably speculative at this point, similarly to other discussions of this topic \cite{carrbook,finebook}. The aim is to stimulate thinking and help envisage general mechanisms involved in setting fundamental constants.

\subsection{Tuning}
\label{tuning}

Regardless of what the actual bounds on fundamental constants from bio-friendly viscosity and diffusion turn out to be, we can ask an interesting question related to fine tuning.

As discussed in the Introduction, fine tuning of fundamental constants was mostly discussed in relation to synthesis of heavy nuclei using theories involved in cosmology, particle physics and astronomy \cite{barrow,hoganreview,adamsreview,uzanreview,carrbook,carr-rees}. According to these theories, some fundamental constants need to be very finely tuned to produce heavy elements. This insight was then used to develop the anthropic argument. In this paper, I extended the length scale by some 15 orders of magnitude from nuclei to observers and found that there are different and additional constraints on fundamental constants coming from bio-friendly (observer-friendly) viscosity and diffusion.

In Section \ref{alphabeta}, I showed that it is possible to produce heavy elements in stars but have a planet where all liquids have very high viscosity due to large $\eta_{min}$ in \eqref{fixed}, for example that of tar or higher and whether observers may not emerge. I noted that the additional tuning due to bio-friendly viscosity is not needed for the generation of heavy nuclei and is therefore {\it redundant} for heavy nuclei. This is relevant to important questions asked previously: (a) can we understand the currently observed values of fundamental constants on the basis of a theory more fundamental than we currently have \cite{barrow,barrow1,weinberg}? and (b) how were these constants tuned \cite{carrbook,finebook}?

One possibility is that fundamental constants were tuned (attained their current values) once. As mentioned earlier, this would involve redundancy involving vast differences between different processes (generating heavy nuclei and bio-friendly viscosity and diffusion in living organisms) extending to 15 orders of magnitude in terms of size and similarly large differences in terms of energy.

If this redundancy is to be avoided, we can conjecture that multiple independent tunings were involved. This includes tuning fundamental constants to produce heavy nuclei and additional tunings needed for other observed sustainable structures to emerge. This conjecture of multiple tuning suggests a similarity to biological evolution where functionally similar traits, such as the different optical nerve connections in human and octopi, were acquired independently. If the analogy is between acquiring a new trait and one act of tuning fundamental constants leading to a new set of these constants, then an organism as a system with multiple separately acquired traits is analogous to the set of observed fundamental constants produced as a result of multiple tunings. In that case, there is no long-standing problem of explaining the observed values of fundamental constants \cite{barrow,weinberg}: currently observed constants should be no more surprising than, for example, the traits acquired by the human eye during biological evolution and the number of these traits, for they are simply the result of adaptive changes. I discuss this in more detail in the next section.

This raises the question of what is the mechanism propagating bio-friendly constants and disfavouring the unfriendly ones during multiple tunings. Some of these mechanisms were discussed. For example, Smolin proposes the cosmological natural selection mechanism where black holes create new universes inheriting fundamental constants close to those in parent universes \cite{smolin} (this concerns fine tuning to produce heavy nuclei only but not other tunings such as those coming from bio-friendly viscosity and diffusion and favouring observers). Harrison discussed the idea that high-intelligent life can create new universes, propagating the fundamental constants further \cite{harrison}. Davies proposes a causal link between laws and product states where physics and biology co-evolve and bio-friendly laws are generated in an evolutionary sense, forming a self-supporting loop \cite{davies}. These proposals meet the challenge of the absence of evidence supporting these processes. Similarly, there is no evidence that life and intelligent life have an effect upon the Universe in the large \cite{barrow1}.

As mentioned in Section \ref{alphabeta}, different viscosity due to different fundamental constants affects not only life functions but also other processes in the cosmic evolution including star formation \cite{turbulence}. The conjecture of multiple tunings does not specify where and when in cosmic history these variations were taking place. There is some evidence that fundamental constants were changing over the large part of the age of the Universe (see, e.g., Refs. \cite{variable1,variable2}). Although this was not upheld in other experiments \cite{grandest,followon,barrowsciadv,followonbeta,betareview}, there are indications of temporal variation of $\alpha$ in other work \cite{seto-variation,lesz}. There is also evidence for spatial variation of $\alpha$ \cite{spatialalpha,spatialalpha1} (see Refs. \cite{chiba,schellekens,martinsreview} for review). Several theoretical models were proposed where fundamental constants vary (see, e.g., Ref. \cite{barrow-var,uzan1,martinsreview,betareview,var-theory}).

\subsection{Evolution}

We should probably take results from evolution involving biology and biochemistry seriously because (a) their principles are well understood and tested, leading to the view that ``nothing in biology makes sense except in the light of evolution'' \cite{dobzhansky} and (b) they involve mechanisms linking inanimate matter to life and observers who are important in the anthropic argument. Below I conjecture that fundamental constants may be related to an evolutionary mechanism. Before discussing this mechanism, I make two remarks.

First, let's recall what we started this paper with and consider a relationship between physical reality, matter or fields, and our mathematical models describing them. A sense in which the two exist independently was discussed: there is a multitude of possible mathematical models (structures) and there is a physical reality corresponding to each model \cite{tegmark}. My view discussed below is more restricted: a mathematical model of a physical object is a practical consequence of its existence and enables us to understand that the object has measurable, consistent and predictable properties. A mathematical model reflects the dynamical rules followed by the object. These rules set the measurable properties of the object, and these properties give the object its distinct existence. The absence of these rules, their inconsistency or violation would imply that we can not specify the properties of this object, making it either non-existent or suggesting that we are dealing with a different system.

Second, let's consider how can fundamental constants be affected by an evolutionary mechanism. Natural selection in biology involves genetic mutations and selection resulting in adaptations to environmental pressure. Let's go back in time and before mutations and selection started and ask how did DNA - an important sustainable mathematical and physical structure of life - emerge? An important insight from biochemistry \cite{lanebook} is that the DNA blocks probably formed in protocells as a result of positive feedback in the metabolic flux. This positive feedback is not just a general idea but is based on specific biochemical processes and reactions: the core metabolism central to life probably started when first catalysts sped up helpful aspects of the metabolic flux in protocells, enabling the conversion of H$_2$ and CO$_2$ into the fabric of new protocells.
The first nucleotides, followed by RNA and DNA, then emerged inside such replicating protocells through positive feedback: protocells with more beneficial chemicals replicated better and passed these chemicals to their daughter cells.

This implies that genetic information emerged from the protocell growth and then helped protocells get better at copying themselves. This enabled protocells to proliferate and hence sustain themselves. Information carried by DNA - the mathematical structure - came into being in the context of growing protocells. ``Meaning emerged with function'', summarises Lane \cite{lanebook}.

This emergence of DNA does not imply that there is an overarching principle that the Universe favor life, cells and observers - the central part of the anthropic argument and SAP in particular. Instead it tells us that nature finds positive feedback pathways to create a sustained and consistent function such as the DNA which, once created, carries an informational mathematical structure to support living cells.

If we view the emergence of information-carrying DNA as a biochemical analogy of mathematical models of particles and fields, we can envisage that parameters in these models, fundamental constants, were tuned as a result of a generally similar evolutionary process. This tuning was multiple as discussed in section \ref{tuning}, and each tuning was analogous to finding one new positive feedback pathway as in the case of emergence of DNA blocks in protocells. A set of fundamental constants stayed if it was helpful to a new stable structure in Figure \ref{visc}.

We don't know enough to discuss details of how the above evolutionary process works in physics (perhaps this is something that the theory of quantum gravity or its successors will aim to achieve). However, the evolutionary mechanism changes the focus of discussion of fine-tuning and fundamental constants and helps address open ``grandest challenge'' questions \cite{grandest}, as follows. Having reviewed long and illustrious history of research related to understanding fundamental constants, Barrow asks: ``will we ever explain the values of constants of Nature?'' \cite{barrow}. For example, can we explain why $\alpha\approx\frac{1}{137}$ or $\beta\approx 1836$? Weinberg notes that we don't know how to calculate fundamental constants in terms of more fundamental constants because we don't know anything more fundamental \cite{weinberg}. In living matter, observed traits such as traits in the human eye are understood on the basis of biological evolution where these traits are a historical legacy of past adaptations. The question why the number of traits is exactly ``X'' rather than ``Y'' is not viewed as meaningful. Nor is this set most optimal (a human eye is less optimised as compared to the octopus eye), as is the case with our Universe which could have been more habitable were some fundamental constants different \cite{adamsreview}. Biochemists may wonder at the well-specified and well-tuned characteristics of the metabolic flux, the basis of life, and ask why didn't the Nature come up with something different and simpler? Having worked out details, biochemists conclude that this is just the result of nature finding the first self-sustained function given the possibilities and constraints of chemistry \cite{lanebook}. An analogy with physics would imply that the observed fundamental constants are the result of nature arriving at sustainable physical structures described by the combination of fundamental theories and fundamental constants,
but the values of these constants may not need to be derived in a more fundamental theory as considered previously \cite{barrow,barrow1,carrbook,finebook,weinberg}. This sustainability may include, for example, idemponents describing a persistent set of generalized points in the algebra of quantum theory \cite{hiley}.


An evolutionary mechanism may help address other questions too, including the perceived ``inelegance'' of the Standard Model with unexplained arbitrary number of generations of quarks and leptons, hierarchy of mass spectra, presence of symmetry in some cases but not in others and so on \cite{zanderighi,uzanbook}. Different views were aired with regard to the role of symmetry. Perhaps the predominant view is that fundamental theories are ruled by symmetry in the aesthetically-driven and potentially unique choice-free grand design \cite{zeesymmetry,weinberg-dreams}. However, Hawking observed that the Standard Model is unlikely to be derivable from a more fundamental theory because is seems accidental (``ugly'' in Hawking's words) rather than being part of the grand design \cite{hawking}. This is consistent with the view that laws of physics can be partly environmental rather than based on a fundamental principle (involving, for example, symmetry) \cite{schellekens}. An evolutionary mechanism, where the emphasis is on function and sustainability, would suggest that symmetry can be helpful to constrain, unite and simplify the process and unhelpful when it hinders versatility and diversity. Consider DNA as an analogue of a mathematical model with its symmetries and fundamental constants setting dynamical rules and instructions for matter and fields. As DNA emerged in protocells due to positive feedback and became helpful in supporting cells, its helix structure understandably acquired some symmetries related to function (the need for the water-soluble phosphate and sugar groups to wrap around the bases and face water on the outside) and local chemistry rules setting interatomic distances, resulting in the helically twisted ladder \cite{dna}. The functions played by sugars and phosphate groups symmetrically wrapping around the bases in DNA are well understood, however questions why the number of bases or sugars and why the resulting DNA symmetry are what they are rather than something else are not viewed as meaningful \cite{dna}. These properties are not considered as fundamental and instead are thought to be the result of what first happened in protocells when the first blocks of DNA emerged \cite{lanebook}. There could probably have been a different outcome (e.g. a different metabolic flux operating in protocells, resulting in different DNA blocks), but what emerged first was helpful and hence stayed.

An analogy with the DNA example above also suggests that not only fundamental constants but also mathematical structures with their symmetries, our physical models, could perhaps also emerge in the process of evolution rather than being part of the grand design from the start. Recall that meaning emerged with function in the context of DNA, the mathematical structure of life. Once emerged, mathematical structures of physical objects stayed to reflect their sustainability, similarly to how DNA helped sustain cells. In terms of mathematical models, what would be an analogue in physics of the first process in protocells where first blocks of DNA emerged due to positive feedback? It might involve a ``protofield'' which we are not aware of experimentally yet. This protofield (perhaps relatable to the quantum gravity theory) would have a property of variability of underlying equations analogous to the variability in evolution. Potential sources of the ability of physical laws to change and transmute were discussed earlier. For example, Bickhard and Campbell \cite{bickhard} and Bickhard \cite{bickhard1} consider how explanations based on variation and selection, due to different sorts of constraints and absorbing conditions, could apply to a wide range of phenomena beyond biology. Wheeler \cite{wheeler} discusses the mutability of physical laws and notes that regularities involved in these laws can ``develop unguided''. This observation was mostly related to the gravitational collapse where space and time are obliterated and so are the physical structures and physical laws whose statement requires space and time. Other sources of variability may include randomness, due to quantum effects, of symmetry-breaking transitions during the Universe cooling resulting in different effective theories potentially emerging \cite{hertog}. Another example is the proposed variability of particle mass and its dependence on time-dependent particle density \cite{massvar1,massvar2}. Finkelstein \cite{finkel} proposes that moving the partition between endosystem and exosystem in a quantum theory results in a variable dynamical law emerging in the endosystem. He also relates evolutionary dynamics to properties of the vacuum consisting of elements capable of reproduction and selection. The mutability of physical laws is also generally consistent with the view that there is a limit to determinism of physical theories and that a theory is an abstraction from a totality that is both qualitatively and quantitatively unlimited \cite{bohmhiley}.

With regard to symmetry, it is interesting to observe that symmetry is appreciated differently depending on the area of science. Symmetry has been helpful to constructing theories and describing experimental data in particle physics \cite{zeesymmetry,weinberg-dreams}. With time, this experience led to the view that aesthetic judgements based on symmetry (both spacetime and internal symmetries) and ``beauty'' in particle physics are effective and advantageous to explaining the data and predicting new experiments. In biology and biochemistry, complexity and the lack of symmetry often accompanying life processes are not viewed as unattractive: scientists in those fields find their objects of study admirable. Perhaps the principles of symmetry and evolution can be combined in a way that gives best understanding of fundamental constants.



\end{document}